\documentclass[12pt]{iopart}

\usepackage{amssymb}

\usepackage{graphicx}

\newcommand{\flabel}[1]{\label{#1}}

\begin{document}

\title{Spurious phase in a model for traffic on a bridge}
\author{David W. Erickson, Gunnar Pruessner, B. Schmittmann and R. K. P. Zia}
\date{today}

\begin{abstract}
We present high-precision Monte Carlo data for the phase diagram of a
two-species driven diffusive system, reminiscent of traffic across a narrow
bridge. Earlier studies reported two phases with broken symmetry; the
existence of one of these has been the subject of some debate. 
We show that the disputed phase disappears for sufficiently
large systems and/or sufficiently low bulk mobility.
\end{abstract}


\address{Center for Stochastic Processes in Science and Engineering and 
Department of Physics, Virginia Tech, Blacksburg, VA 24061-0435,
USA}
\ead{\mailto{daericks@vt.edu}, \mailto{gunnar.pruessner@physics.org},
\mailto{schmittm@vt.edu}, \mailto{rkpzia@vt.edu}}

\submitto{\JPA}


\section{Introduction}
When driven far from thermal equilibrium, interacting many-particle
systems display a wide variety of unexpected behaviours
\cite{SchmittmannZia:1995,Mukamel:2000}. Generic long-range
correlations, new phase transitions and universality classes, and
anomalously fast coarsening are amongst the most striking phenomena. In
contrast to equilibrium systems, our notions of energy-entropy
competition fail so that, even in low spatial dimensions, a plethora of
novel phases and phase transitions have been discovered
\cite{SchmittmannZia:1995,Mukamel:2000,Schuetz:2000}.  In the absence of
a fundamental theoretical framework for nonequilibrium systems, these
transitions are usually identified with the help of Monte Carlo
simulations, and subtle finite-size analyses are essential in order to
establish their nature in the thermodynamic limit. Only in a few cases
can such questions be settled by recourse to exact solutions.

Asymmetric exclusion processes (ASEPs) constitute a particularly simple
class of far-from-equilibrium model systems \cite{Schuetz:2000}. Over
the past two decades, they have emerged as a kind of ``laboratory'' in
which nonequilibrium phase transitions can be studied. ASEPs have also
been invoked to model a variety of physical systems, such as
biological transport,
traffic problems
and gel electrophoresis.
Defined on a regular \emph{one-dimensional} lattice of length $L$, particles and holes
perform nearest-neighbour exchanges biased in a specific direction. If
exchanges in the suppressed direction are completely prohibited, the
process becomes totally asymmetric (TASEP). For periodic boundary
conditions, the TASEP stationary state is uniform; in contrast, if
particles are inserted and extracted at the boundaries with certain
rates ($\alpha$ and $\beta$, respectively), three distinct phases are
observed: a low-density ($\alpha<1/2$, $\alpha <\beta$), a high-density
($\alpha >\beta$, $\beta <1/2$), and a maximal current ($\alpha >1/2$,
$\beta >1/2$) phase
\cite{Krug:1991,SchuetzDomany:1993,DerridaETAL:1993}.

If two species of particles are present, phase transitions can occur
even in periodic systems \cite{EvansETAL:1998,EvansETAL:1998b}.  If the
two species (labelled ``positive'' and ``negative'') are biased in
opposite directions, move into vacant sites (holes) with unit rate, and
exchange with the opposite species with another rate ($\gamma <1$), then
solid clusters of particles separated by nearly empty domains form
easily.  Though this particular model is proved to display only a
homogeneous phase \cite{SandowGodreche:1997}, dramatically different
properties emerge when it is generalized in a seemingly trivial manner - from
$1\times L$ to $2\times L$, modeling a ``\emph{two lane} road'' 
(see \cite{GeorgievSchmittmannZia:2005} and references therein) .

In this Letter, we consider the same biased two-species dynamics, but with open
boundaries. In this ``bridge'' model, positive (negative) particles enter
the system at the left (right) with rate $\alpha$ and exit at the right
(left) with rate $\beta$, reminiscent of traffic crossing a long road
bridge in both directions. While cars move freely on clear stretches, they
slow down for on-coming traffic, reflected in the bulk mobility $\gamma$.
This dynamics is clearly symmetric with respect to charge-parity (CP)
transformations. The first study \cite{EvansETAL:1995} of this model
focused mostly on $\gamma =1$ (i.e., particle-particle and particle-hole
exchanges occur with the same rate in the bulk) and reported four phases.
Two of these phases are characterised by CP-symmetric steady-state particle
densities and currents, while the other two 
spontaneously break the CP-symmetry.
Specifically, choosing $\alpha =1$ and
lowering $\beta$, the system undergoes a transition from a symmetric (the
so-called ``power-law'') phase with maximal current, equal particle
densities and no holes to a low-density symmetric (\emph{ld-S}) phase where
the particle densities are still equal but below $1/2$. As $\beta$ is
lowered further, spontaneous symmetry breaking occurs: While both
densities remain below $1/2$, the two species are associated with
different currents and densities. At even lower $\beta$, this
``low-density low-density'' (\emph{ld/ld}) phase undergoes a final
transition into a ``high-density low-density'' (\emph{hd/ld}) phase in
which one of the particle densities exceeds $1/2$. While the existence of
the two symmetric phases is supported by an exact solution for $\beta =1$,
evidence for the two asymmetric phases is restricted to a mean-field theory
and simulation data for fairly small systems. The existence of the 
(\emph{ld/ld})-phase was subsequently disputed \cite{ArndtHeinzelRittenberg:1998}, 
based predominantly on simulations of a suitably defined free energy
functional which can be extracted from particle density histograms. Arguing
that this functional is not very sensitive to the subtle differences between
the (\emph{ld/ld})- and (\emph{hd/ld})-phases, more accurate data for
particle density histograms were analysed in \cite{ClincyEvansMukamel:2001},
providing further evidence for the (\emph{ld/ld})-phase albeit with some
unusual finite-size properties.

In this Letter, we investigate this model further, performing high-precision Monte
Carlo (MC) simulations for much larger system sizes and an expanded range of 
$\gamma$, with $0<\gamma \leq 1$, 
in an attempt to clarify the
nature of the (\emph{ld/ld})-phase. We present very strong numerical
evidence that only a single symmetry breaking transition persists in the
large $L$ or small $\gamma$ limit, in the sense that the range in $\beta$
over which the (\emph{ld/ld})-phase is observed narrows until it becomes
unresolvable. As in the periodic case \cite{GeorgievSchmittmannZia:2005}, 
the behaviour of low-dimensional
systems can easily be misextrapolated unless very careful finite-size
analyses are performed. Our findings 
confirm how subtle finite-size effects can be when studying driven
diffusive systems. The importance of exploring larger and larger systems
cannot be overemphasised.
Since a reliable extrapolation of MC data is essential for the
analysis of many nonequilibrium phenomena, the understanding of such
peculiar finite-size effects may have repercussions well beyond these simple
models.

This Letter is organised as follows. We first introduce the model and
our simulation method, then summarise our data, and conclude with a
brief summary. More details will be reported in \cite{EricksonETAL:TBD}.

\section{Model and methods}

The model is defined on a one-dimensional lattice of length $L$. An
occupation variable $s_{i}=(\cdot )_{i}$ is assigned to each site 
$i=1,...,L$, indicating its state as empty, $(0)_{i}$, or occupied by a positive, 
$(+)_{i}$, or negative, $(-)_{i}$, particle. Positive particles
enter the system on the left with rate $\alpha$ and move to the right with
rate $1$ if their neighbour is a hole, or exchange places with a negative
particle with rate $\gamma$. At the right edge, they leave the system with
rate $\beta$. Negative particles enter the system from the right and exit
at the left, in a CP-invariant fashion. In detail, 
\begin{equation}  \label{processes}
\begin{tabular}[t]{lllllll}
& $(0)_{1}$ & $\to$ & $(+)_{1}$ &  & $\quad \textrm{with rate } \alpha$ & 
\qquad (a) \\
& $(0)_{N}$ & $\to$ & $(-)_{N}$ &  & $\quad \textrm{with rate } \alpha$ & 
\qquad (b) \\
& $(+)_{N}$ & $\to$ & $(0)_{N}$ &  & $\quad \textrm{with rate } \beta$ & 
\qquad (c) \\
& $(-)_{1}$ & $\to$ & $(0)_{1}$ &  & $\quad \textrm{with rate } \beta$ & 
\qquad (d) \\
$(+)_{i}$ & $(-)_{i+1}$ & $\to$ & $(-)_{i}$ & $(+)_{i+1}$ & $\quad
\textrm{with rate } \gamma$ & \qquad (e) \\
$(+)_{i}$ & $(0)_{i+1}$ & $\to$ & $(0)_{i}$ & $(+)_{i+1}$ & $\quad
\textrm{with rate } 1$ & \qquad (f) \\
$(0)_{i}$ & $(-)_{i+1}$ & $\to$ & $(-)_{i}$ & $(0)_{i+1}$ & $\quad
\textrm{with rate } 1$ & \qquad (g)
\end{tabular}
\end{equation}

To implement this dynamics, we maintain a list of possible moves, draw
randomly from that list and update according to the rates in
\eref{processes}.  If the length of the list is $l$, continuous time,
measured in units of MC steps (MCS), is incremented by $1/l$ whenever a
move is attempted. Taking the average of an observable then requires a
weighting by $1/l$. The performance of the code can be greatly improved
if we note that the particle density changes only at the boundaries,
\eref{processes}(a)--(d) \cite{JensenPruessner:2003b}.

All simulations are performed at $\alpha =1$. Starting from an initially
randomly half-filled system, we discard at least $10^{9}$ MCS to avoid
transients, and then gather data for about $4\times 10^{9}$ MCS, unless
otherwise noted. Smaller values of $\gamma$ require longer runs, up to
$2.5\times 10^{10}$ MCS, to gain a clear picture. The random number
generator is the ``Mersenne twister'' \cite{MatsumotoNishimura:1998b}.

We generate particle density histograms, $P_{L}(\rho_{-},\rho_{+})$, by
continuously monitoring the instantaneous densities of positive and
negative particles, $\rho_{\pm}\equiv (2L)^{-1}\sum_{i}\left[ \left|
s_{i}\right| \pm s_{i}\right]$, in steady state. The associated
configurational averages are denoted by $\langle \rho_{+}\rangle$ and
$\langle \rho_{-}\rangle$. 

In steady state, the probability of a particular configuration of a
\emph{finite} system is rigorously symmetric under CP, so that
$P_{L}(\rho_{-},\rho_{+})=P_{L}(\rho_{+},\rho_{-})$ and $\langle
\rho_{-}\rangle =\langle \rho_{+}\rangle$.  We exploit this symmetry in
order to reduce noise in our statistical data, by considering the
``folded'' histogram $P_L(\rho_{-},\rho_{+})+P_L(\rho_{+},\rho_{-})$.  In
certain sectors of parameter space, CP-symmetry is spontaneously broken
\cite{EvansETAL:1995}, in the sense that larger systems are ``trapped''
for times $\propto \exp (const\times L)$ in particular regions of phase
space with asymmetric particle densities and currents.

Even for short switching times, the particle density histograms
$P_L(\rho_{-},\rho_{+})$ carry by far the clearest signature of the
transitions \cite{ArndtHeinzelRittenberg:1998,ClincyEvansMukamel:2001}.
A single peak on the diagonal indicates a symmetric phase, while a
double peak with two off-diagonal maxima signals a symmetry-broken
phase.  The transition between the two asymmetric phases is marked by
histograms with two long ridges, one running close and nearly parallel
to the $\rho_{-}$-axis, and the other close to the $\rho_{+}$-axis. At
the transition, the global maximum suddenly shifts from the near
(\emph{ld/ld}) to the far (\emph{hd/ld}) end of each ridge, like water
sloshing in a narrow channel. Going beyond earlier studies, we
\emph{quantify} the histogram features by measuring the observables
$\rho_{\textrm{{\tiny min}}}$ and $\rho_{\textrm{{\tiny maj}}}$,
denoting the density of the minority ($\rho_{\textrm{{\tiny min}}}$) and
majority ($\rho_{\textrm{{\tiny maj}}}$) species at the \emph{global}
maximum of $P_L(\rho_{-},\rho_{+})$. The statistical error of the peak
position can be determined from its scatter in independent runs. This is
computationally very costly, so that the error bars shown for $L=500$ in
\fref{gap_gamma1} are extrapolated from a few runs at representative
parameter values. They are typically of the size of the symbols.

\section{Results}
\begin{figure}[t]
\begin{center}\scalebox{0.42}{\includegraphics*{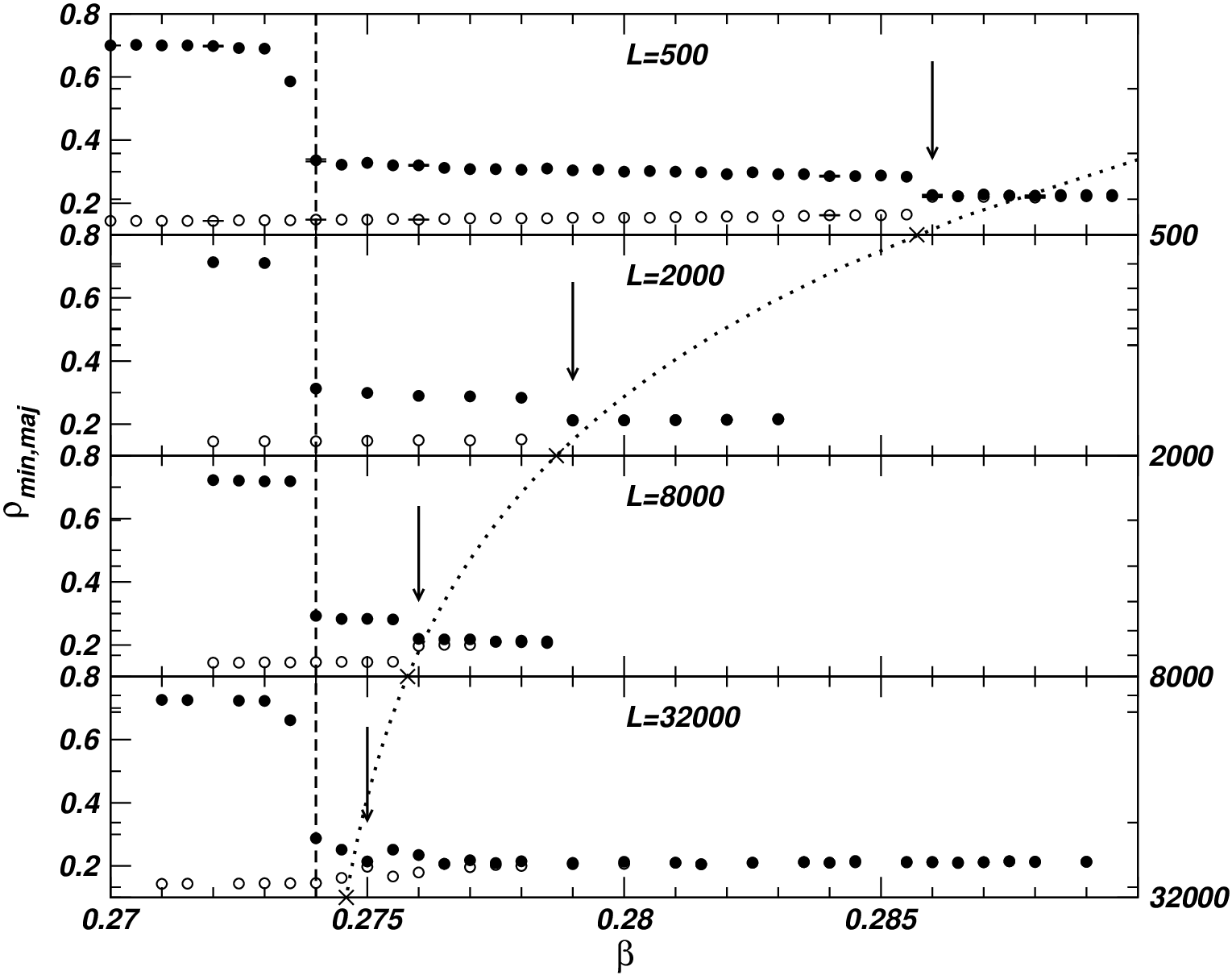}}\end{center}
\caption{
\flabel{gap_gamma1}
The position of the global maximum in the histogram
$P_L(\rho_{-},\rho_{+})$ as a function of $\beta$ at $\gamma=1$ for
different system sizes $L$. The filled (open) symbols show the density
of the majority (minority) species, $\rho_{\textrm{\tiny maj}}$ and
$\rho_{\textrm{\tiny min}}$, respectively. The smallest value of $\beta$
in the symmetric phase is marked by an arrow, indicating the onset of
the (\emph{ld-S})-(\emph{ld/ld}) transition. It moves significantly with
$L$. In contrast, the (\emph{ld/ld})-(\emph{hd/ld}) transition, marked
by a dashed line through all four panels (smallest $\beta$ in
(\emph{ld/ld})), stays fixed.  The dotted line (right ordinate) is a
power law fit
of the position of the (\emph{ld-S})-(\emph{ld/ld}) transition, $L$ vs.
$\beta_2(L)$, the crosses marking the suggested transition point for
each system size shown.}
\end{figure}

We first consider the $L$-dependence of the phase diagram, for $\gamma
=1$.  \Fref{gap_gamma1} shows $\rho_{\textrm{{\tiny min}}}$ and
$\rho_{\textrm{{\tiny maj}}}$ for different system sizes
$L=500,2000,8000,32000$ at $\gamma =1$.  Starting at low $\beta$, each
system size displays two branches, the upper (lower) being associated
with $\rho_{\textrm{{\tiny maj}}}$ ($\rho_{\textrm{{\tiny min}}}$). 
The (\emph{hd/ld})-(\emph{ld/ld}) transition, associated with the abrupt
drop in $\rho_{\textrm{\tiny maj}}$, occurs at a nearly $L$-independent
value of $\beta_1 \simeq 0.2735$.
Even though the difference
$\rho_{\textrm{{\tiny maj}}}-\rho_{\textrm{{\tiny min}}}$ is
significantly reduced at the transition, it remains nonzero above
$\beta_{1}$, indicating \emph{asymmetric} phases on both sides.
Remarkably, only $\rho_{\textrm{{\tiny maj}}}$ signals the transition
clearly while $\rho_{\textrm{{\tiny min}}}$ is quite smooth. As we
increase $\beta$ further, we encounter the (\emph{ld/ld})-(\emph{ld-S})
transition where $\rho_{\textrm{{\tiny maj}}}-\rho_{\textrm{{\tiny
min}}}$ vanishes, up to statistical fluctuations. Labelling the
transition point $\beta_{2}(L)$, it is obvious that $\beta_{2}$ depends
strongly on $L$. Our data for the \emph{smallest }system size ($500$)
confirm the results of \cite{ClincyEvansMukamel:2001}, i.e.,
$\beta_{1}\simeq 0.274$ and $\beta_{2}(500)\simeq 0.284$. However, the
rest of our data show clearly that $\beta_{2}(L)$ is
monotonically \emph{decreasing} with $L$. Although the data for
$L=32000$ is quite noisy (with runs of only $2.2 \times 10^{8}$ MCS
overall), one can still
roughly locate the second transition. 
In particular, we expect $\beta _2(L)$ to be just below
the smallest $\beta$ for which 
$\rho_{\textrm{{\tiny maj}}}=\rho_{\textrm{{\tiny min}}}$  
(indicated by arrows in \fref{gap_gamma1}). From these
estimates of $\beta_2(L)$, we found that the data are remarkably consistent
with the power law $\beta_2(L)-\beta_1\propto L^{-0.63}$. The dotted line
in \fref{gap_gamma1} shows such a fit. Barring further
surprises at even larger values of $L$, we take this as a strong
indication that the (\emph{ld/ld})-phase does not persist in the
$L\rightarrow \infty$ limit. As a further test, we investigated the
maximum of $P_{L}(\rho_{-},\rho_{+})$,
$\max_{\rho_{-},\rho_{+}}\{P_{L}(\rho_{-},\rho_{+}) |
\rho_{-}-\rho_{+}\}$, along diagonals defined by constant
$\rho_{-}-\rho_{+}$ \cite{ClincyEvansMukamel:2001}.
Examples for these ``ridge profiles'' or ``silhouettes'' are shown in
\fref{phase_diagram}.  For lower values of $L$ and \emph{fixed} $\beta$
with, say,  $\beta_{1} < \beta < \beta_{2}(500)$, the ratio of peak to
saddle height of this function increases with $L$, which appears to
support the presence of the (\emph{ld/ld})-phase
(see figure 10 in \cite{ClincyEvansMukamel:2001}). 
Yet for sufficiently large $L$, this
ratio decreases again and eventually drops to $1$ as soon as $\beta_2$,
the position of the (\emph{ld/ld})-(\emph{ld-S}) transition, has moved
below the $\beta$ considered. We expect to see similar behaviour for any
value of $\beta$ between $0.2735$ and $0.284$.

\begin{figure}[t]
\begin{center}\scalebox{0.42}{\includegraphics*{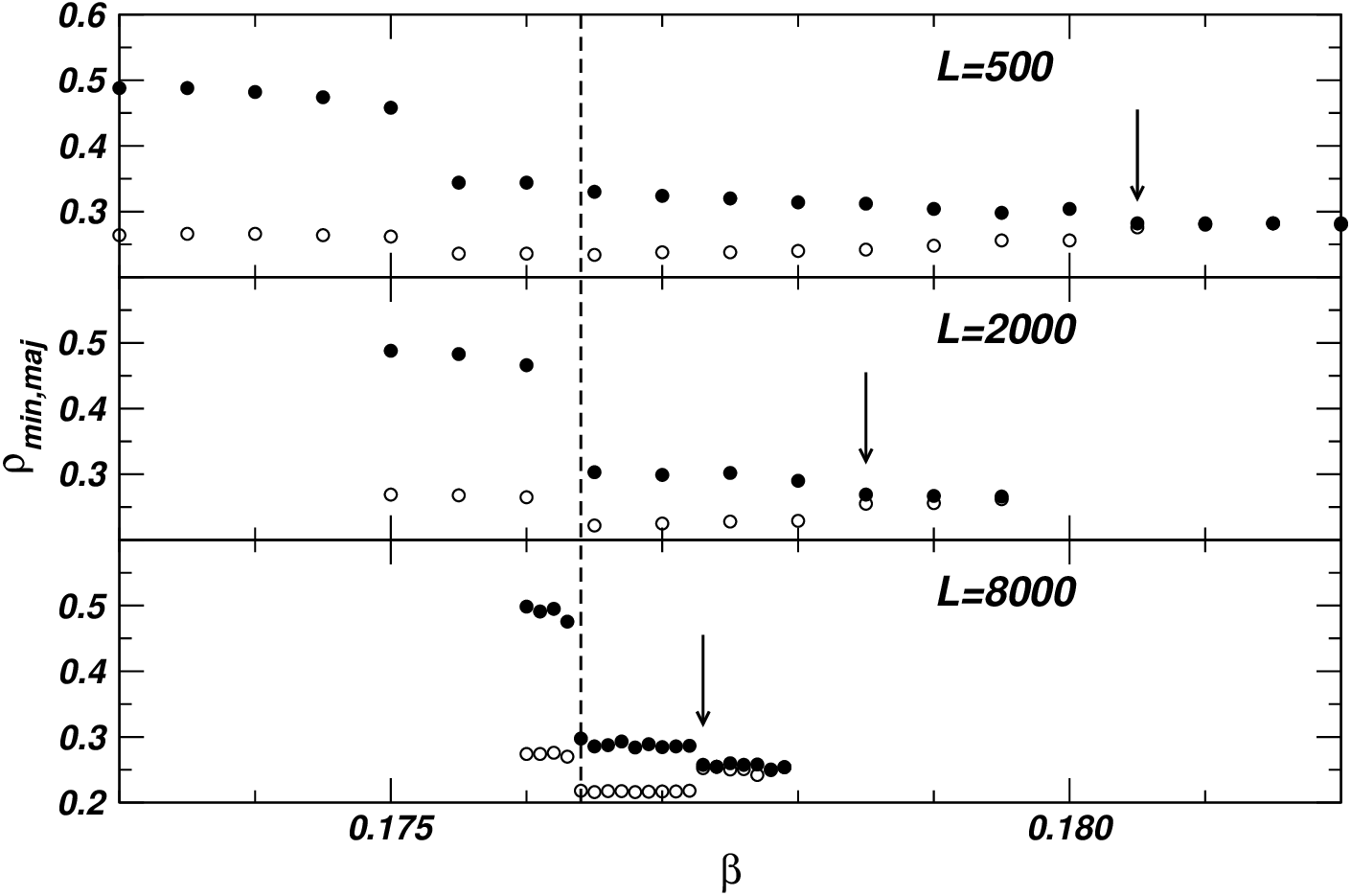}}\end{center}
\caption{
\flabel{gap_gamma0.4}
As \fref{gap_gamma1}, the position of the global maximum in the
histogram $P_L(\rho_{-},\rho_{+})$ as a function of $\beta$ for different
system sizes $L$, but now for $\gamma=0.4$. Filled (open) symbols denote
the majority (minority) density,  $\rho_{\textrm{\tiny maj}}$
($\rho_{\textrm{\tiny min}}$).  The position of the transition
(\emph{ld-S})-(\emph{ld/ld}), marked by arrows, moves towards the
(\emph{ld/ld})-(\emph{hd/ld}) transition.  The latter moves slightly
towards larger $\beta$ with $L$, the dashed line marking the
largest $\beta$ in (\emph{ld/ld}) for the largest $L$.}
\end{figure}

These data provide a clear picture of the $\gamma =1$ model. We now turn
to an expanded parameter space by considering smaller values of
$\gamma$, reflecting congestions on the bridge. Based on extensive
simulations (up to $5\times 10^{9}$ and $2\times 10^{10}$ MCS for the
transient and statistics, respectively), the transitions can be
investigated for $\gamma$ as low as $0.4$. \Fref{gap_gamma0.4} shows,
similarly to \fref{gap_gamma1}, the behaviour of minority and majority
species. Our key observation is that the width of the
(\emph{ld/ld})-phase is even further reduced. For $\gamma =0.4$ and
$L=500$, it is already confined to $0.175\leq \beta \leq 0.180$,
shrinking further to $0.1763\leq \beta \leq 0.1773$ for $L=8000$, with
errors of at most $\pm 0.0005$. Unless very finely grained steps in
$\beta$ are simulated, it is easy to miss this phase entirely. It is
also apparent that both transitions move to smaller values of $\beta$ as
$\gamma$ decreases; however, $\beta_{2}$ decreases faster so that the
width of the (\emph{ld/ld})-phase shrinks as $\gamma$ is lowered. Below
$\gamma =0.4$ the data becomes very noisy so that only moderately large
systems can be considered.  Here, the histograms develop only short
ridges, very close to the diagonal $\rho_{+}=\rho_{-}$. For small
$\beta$, the global maximum is at the far end of the ridge,
corresponding to the (\emph{hd/ld})-phase. As $\beta$ increases, the
ridge shortens and moves closer to the diagonal. During this reshaping,
it becomes virtually impossible to resolve whether the global maximum
actually moves to the near end of the ridge -- which would be the
signature of the (\emph{ld/ld})-phase -- or whether it just immediately
appears on the diagonal, signalling the (\emph{ld-S})-phase. Also, the
low-$\beta$ phase is no longer characterised by having
$\rho_{\textrm{{\tiny maj}}}>1/2$; instead, both $\rho_{\textrm{{\tiny
maj}}}$ and $\rho_{\textrm{{\tiny min}}}$ are below $1/2$; in this
sense, the terminology (\emph{hd/ld}) is somewhat of a misnomer. To
conclude, the lowest $\gamma$ for which we still identify the
(\emph{ld/ld})-phase with reasonable certainty is $\gamma =0.4$ for
$L=500$. The phase diagram in \fref{phase_diagram} summarises our
findings, showing the position of the (\emph{hd/ld})-(\emph{ld/ld}) and
(\emph{ld/ld})-(\emph{ld-S}) transitions in the $\beta, \gamma$-plane
for $L=500$. Below $\gamma =0.4$, we draw only a single phase boundary,
separating a symmetry-broken phase from a symmetric one. Remarkably, the
whole length of the (\emph{hd/ld})-phase boundary can be well approximated
by the phenomenological formula $1-\sqrt{1-\beta
/\beta _1}$ with $\beta _1=0.2735$. If a theoretical framework to predict
this formula were developed, it would undoubtedly provide significant
insight into the complex behaviour of this simple model.

\begin{figure}[t]
\begin{center}\scalebox{0.37}{\includegraphics*{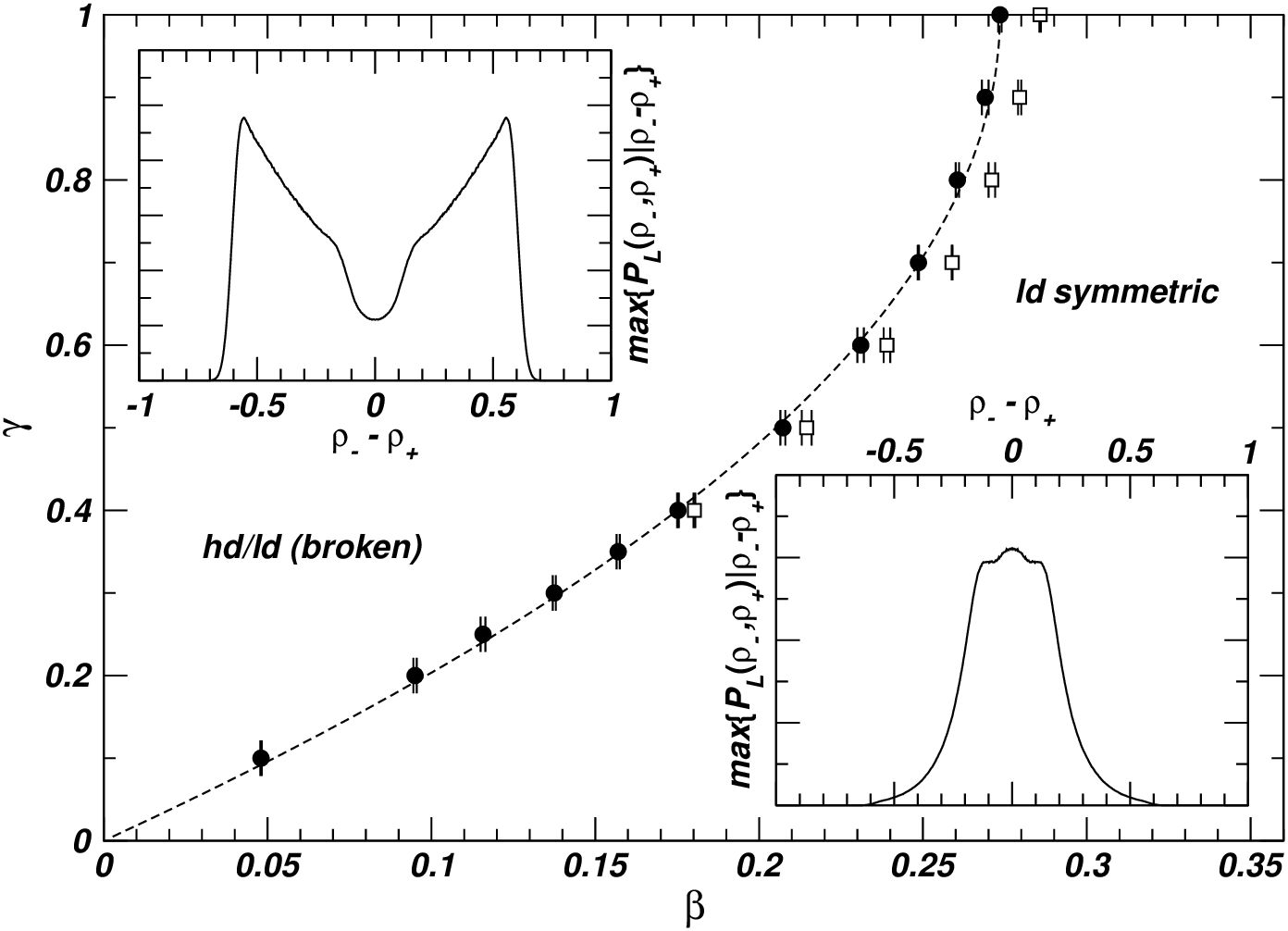}}\end{center}
\caption{
\flabel{phase_diagram}
Phase diagram in $\beta$, $\gamma$ space for the bridge model at
$L=500$.  The closed symbols show the transition
(\emph{hd/ld})-(\emph{ld/ld}), the open symbols the transition
(\emph{ld-S})-(\emph{ld/ld}), which is expected to merge with the
former. The error bars indicate the estimated range of the transition.
The dashed line is a rough fit of the first transition to
$1-\sqrt{1-\beta /\beta_{1}}$ with $\beta_{1}=0.2735$.  The two insets
show typical ``ridge profiles'', i.e., the maximum of
$P_L(\rho_{-},\rho_{+})$ at given $\rho_--\rho_+$, for the two major
phases at $\beta=0.271$ and $\beta=0.287$, both for $\gamma=1$ and
$L=500$.}
\end{figure}

\section{Conclusions}
Using extensive simulations for much larger system sizes than previously
studied, we performed a finite-size analysis for a simple two-species
asymmetric exclusion process with open boundaries. The model is
reminiscent of traffic crossing a narrow bridge in both directions.
Extending earlier work, we considered the effect of having a small
passing probability, $\gamma$, which is the rate for two particles to
exchange places. In smaller systems with $\gamma \lesssim 1$, 
we confirm the existence of four phases. However, as $L$
increases or $\gamma$ decreases, the (\emph{ld/ld})-sector shrinks until
it can no longer be resolved for $\gamma \leq 0.4$ and $L\geq 500$.
This suggests strongly that the (\emph{ld/ld})-phase is a finite-size
effect and that there is only one symmetry-breaking transition in
this model. While mean-field theory predicts otherwise, it is often
unreliable in low dimensions. 
Our investigation highlights again
the importance of using very large systems in numerical studies of driven
diffusive systems. Moreover, it suggests that scanning a wider range of
control parameters may already reveal the same subtleties in smaller systems.
Until a reliable analytic approach for this model is found, we believe that our 
study provides the best insight so far into the issues raised in earlier works
\cite{ArndtHeinzelRittenberg:1998,ClincyEvansMukamel:2001}.

\emph{Acknowledgements.} We thank Ivan Georgiev and Gunther Sch\"{u}tz
for fruitful discussions, and the Virginia Tech Terascale Computing
Facility and the Mathematics Department at Imperial College London, in
particular Andy Thomas, for technical support. This work is partially
supported by NSF through DMR-0308548 and DMR-0414122. GP acknowledges
the Alexander von Humboldt foundation for their support.

\section*{References}
\bibliography{articles,books}
\end{document}